\documentclass[pdftex,aps,prl,amsmath,amsfonts,amssymb,superscriptaddress,twocolumn,10pt]{revtex4}
\renewcommand{\vec}[1]{\mathbf{#1}}
\usepackage[pdftex]{graphicx}
\usepackage{multirow}
\usepackage{wasysym}
\usepackage{setspace}
\usepackage{color}

\renewcommand{\Re}{\operatorname{Re}}
\renewcommand{\Im}{\operatorname{Im}}

\newcommand{\figref}[1]{Fig.~\ref{fig:#1}}
\newcommand{\Figref}[1]{Figure~\ref{fig:#1}}

\renewcommand{\eqref}[1]{Eq.~(\ref{eq:#1})}
\newcommand{\Eqref}[1]{Equation~(\ref{eq:#1})}

\def\a{s}
\def\b{s}
\newcommand{\add}[1]{\if\a\b{{\color{red} #1}}\else{#1}\fi}
\newcommand{\comm}[1]{\if\a\b{{\color{blue}\{\small \sc #1\}}}\else{}\fi}
\newcommand{\del}[1]{{\if\a\b{{\color{magenta}[[#1]]}}\else{}\fi}}
\newcommand{\sgj}[1]{\if\a\b{{\color{red}\{\small \sc #1\}}}\else{}\fi}

\begin{document}

\title{Ingredients of a Casimir Analog Computer}

\author{Alejandro~W. Rodriguez}
\affiliation{Department of Physics,
Massachusetts Institute of Technology, Cambridge, MA 02139}
\author{Alexander~P. McCauley}
\affiliation{Department of Physics,
Massachusetts Institute of Technology, Cambridge, MA 02139}
\author{John D. Joannopoulos}
\affiliation{Department of Physics,
Massachusetts Institute of Technology, Cambridge, MA 02139}
\author{Steven G. Johnson}
\affiliation{Department of Mathematics,
Massachusetts Institute of Technology, Cambridge, MA 02139}

\begin{abstract}
  We present the basic ingredients of a technique to compute quantum
  Casimir forces at micrometer scales using antenna measurements at
  tabletop (e.g.~centimeter) scales, forming a type of analog computer
  for the Casimir force.  This technique relies on a correspondence
  that we derive between the contour integration of the Casimir force
  in the complex frequency plane and the electromagnetic response of a
  physical dissipative medium in a finite real-frequency bandwidth.
\end{abstract}

\maketitle

\begin{figure}[tp]
\includegraphics[width=1.0\columnwidth]{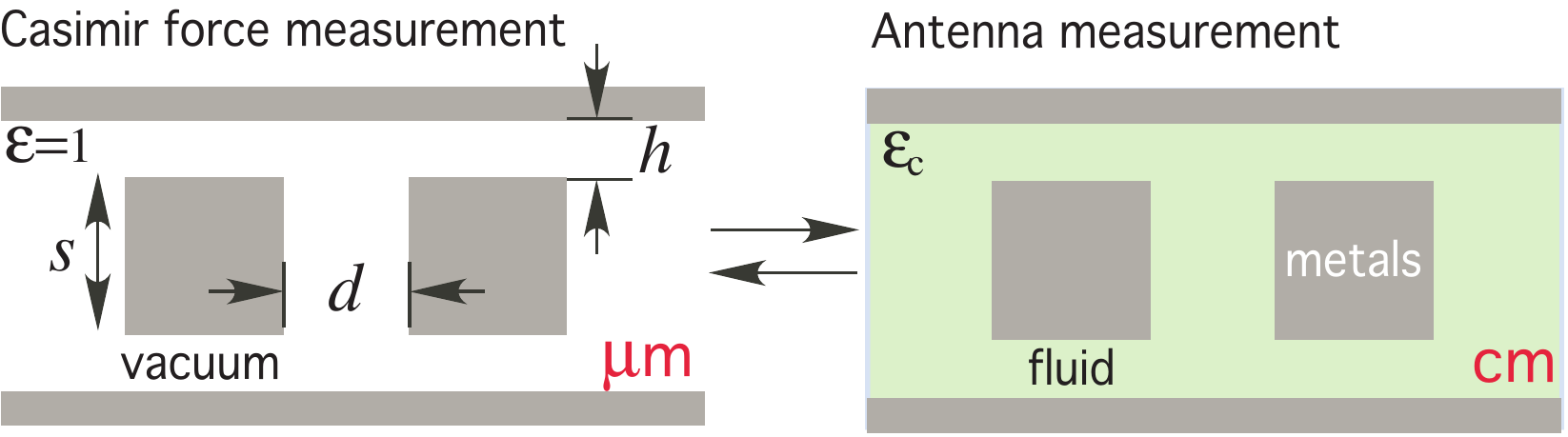}
\caption{Schematic illustration of correspondence between observations
of Casimir forces at $\mu$m lengthscales and antenna measurements of
the electromagnetic response functions at tabletop lengthscales
(e.g. cm).}
\label{fig:schematic}
\end{figure}

Casimir forces arise due to quantum fluctuations of the
electromagnetic field~\cite{casimir} and can play a significant role
in the physics of neutral, macroscopic bodies at micrometer
separations, such as in new generations of microelectronic mechanical
systems (MEMS)~\cite{Serry98, hochan1}.  These forces have previously
been studied both in delicate experiments at micron and sub-micron
lengthscales~\cite{bordag01} and also in theoretical calculations that
are only recently becoming feasible for complex non-planar
geometries~\cite{emig06, Rodriguez07:PRA}. Here, we propose a third
alternative by deriving an equivalence between quantum fluctuations
and the classical electromagnetic response in bodies separated by a
conducting fluid, as illustrated in \figref{schematic}.  Using this
equivalence, we propose the possibility of experimental $S$-matrix
measurements for microwave antennas in centimeter-scale models that
indirectly yield the Casimir force between micron-scale objects.  Such
a centimeter-scale model is not a Casimir ``simulator,'' in that one
is not measuring forces, but rather a quantity that is mathematical
related to the micron-scale Casimir force---in this sense, it is a
kind of \emph{analog computer}. We believe that this mathematical
equivalence between disparate quantum and classical systems reveals
new opportunities for the experimental and theoretical study of
Casimir interactions.

In the following letter, we first review a well-known formulation of
the Casimir force in terms of the classical Green's function (GF) via
the electromagnetic stress tensor (ST) and the fluctuation-dissipation
theorem~\cite{Lifshitz80}.  Although this formulation is normally
expressed for either real or imaginary frequencies, we consider the
general complex-frequency ($\omega$) plane.  We then show that the
mapping to complex frequency is equivalent to a real-frequency GF with
a transformed electromagnetic medium $\varepsilon_c$.  From this point
of view, however, it turns out that only certain contours in the
complex-frequency plane correspond to physically realizable
$\varepsilon_c$, and in particular the real and imaginary frequency
axes are unsuitable.  Instead, we identify another contour and show
its equivalence to a conventional conducting medium, demonstrate that
the response of such a medium yields the correct Casimir force in
nontrivial geometries, and consider the implications and possible
materials for practical experiments.  The key point is that, once the
Casimir force is expressed in terms of the response of a realizable
medium over a reasonably narrow bandwidth, the scale-invariance of
Maxwell's equations permits this response to be measured at any
desired lengthscale, e.g. in a tabletop microwave experiment.

The Casimir force can be expressed as an integral of the mean
electromagnetic ST over all frequencies~\cite{Lifshitz80}.  The mean
ST is determined simply from the classical GF (the fields in response
to current sources at a fixed frequency), thanks to the
fluctuation-dissipation theorem.  It turns out, however, that this
frequency integral is badly behaved from the perspective of numerical
calculations (or experiments, below).  Fortunately, because the
integrand is analytic, one can deform the integration contour into the
complex-frequency $\omega$ plane.

More generally, given an arbitrary contour $\omega(\xi)$ (for
convenience below, we parameterize the contour by a real $\xi$),
the force in the $i$-th coordinate direction is given by:
\begin{equation}
    F_i = \Im \int_0^\infty d\xi \frac{d\omega}{d\xi}
    \oiint_{\mathrm{surface}} \sum_j \langle T_{ij}(\vec{r},\omega)
    \rangle \, dS_j \, .
\end{equation}
The standard Wick rotation corresponds to the particular choice
$\omega(\xi) = i\xi$ and yields a smooth and rapidly decaying
integrand~\cite{Rodriguez07:PRA}.  The mean ST $\langle T_{ij}
\rangle$ is related to the electric ($\vec{E}$) and magnetic
($\vec{H}$) field correlation functions by the standard equation
(assuming non-magnetic materials, $\mu = 1$, for simplicity):
\begin{multline}
\left\langle T_{ij} (\vec{r},\omega) \right\rangle = \left\langle
H_{i}(\vec{r})\,H_{j}(\vec{r})\right\rangle
-\frac{1}{2}\delta_{ij}\sum_k\left\langle
H_{k}(\vec{r})\,H_{k}(\vec{r})\right\rangle \\ +
\varepsilon(\vec{r},\omega) \Big[ \left\langle
  E_{i}(\vec{r})\,E_{j}(\vec{r})\right\rangle
  -\frac{1}{2}\delta_{ij}\sum_k \left\langle
  E_k(\vec{r})\,E_k(\vec{r})\right\rangle \Big] \,.
\end{multline}
The field correlation functions are, in turn, related to the
frequency-domain \emph{classical} photon GF,
$G_{ij}(\omega;\vec{r},\vec{r}')$, by the fluctuation-dissipation
theorem:
\begin{gather}
\label{eq:ecorr}
\left\langle E_{i}(\vec{r}) E_{j}(\vec{r}')\right\rangle =
\frac{\hbar}{\pi} \omega^2 G_{ij}(\omega;\vec{r},\vec{r}')
\\ \left\langle H_{i}(\vec{r}) H_{j}(\vec{r}')\right\rangle
=-\frac{\hbar}{\pi} (\nabla\times)_{i\ell}(\nabla'\times)_{jm} G_{\ell
  m}(\omega;\vec{r}, \vec{r}') \, ,
\end{gather}
where $G_{ij}$ satisfies Maxwell's equations:
\begin{equation}
\left[\nabla\times\nabla\times{}-\varepsilon(\vec{r},\omega)\omega^2\right]\vec{G}_{j}(\omega;\vec{r},\vec{r}')=\delta(\vec{r}-\vec{r}')\hat{\vec{e}}_{j}
\label{eq:Green-gen}
\end{equation}
\Eqref{Green-gen} can be solved in a number of ways, for example by a
finite-difference discretization~\cite{Rodriguez07:PRA} or even
analytically in one dimension~\cite{Lifshitz80}.  Of course, the
diagonal ($\vec{r}'=\vec{r}$) part of the GF is formally infinite, but
this singularity is not relevant because its surface integral is zero,
and it is typically removed by some regularization (e.g. by the finite
discretization or by a finite antenna size in the proposed experiments
below).  A crucial step, as mentioned above, is the passage to
imaginary frequencies $\omega(\xi) = i\xi$.  For real frequencies, the
GF is oscillatory, leading to a highly oscillatory ST integrand that
does not decay---even when a regularization (ultraviolet cutoff) is
imposed, integrating a highly oscillatory function over a broad
bandwidth is problematic.  For imaginary frequencies, on the other
hand, the GF is exponentially decaying, due to the operator in
\eqref{Green-gen} becoming positive-definite
($\nabla\times\nabla\times{}+\varepsilon\xi^2$)~\cite{Rodriguez07:PRA},
leading to a decaying non-oscillatory integrand.

\begin{figure}[tp]
\includegraphics[width=0.9\columnwidth]{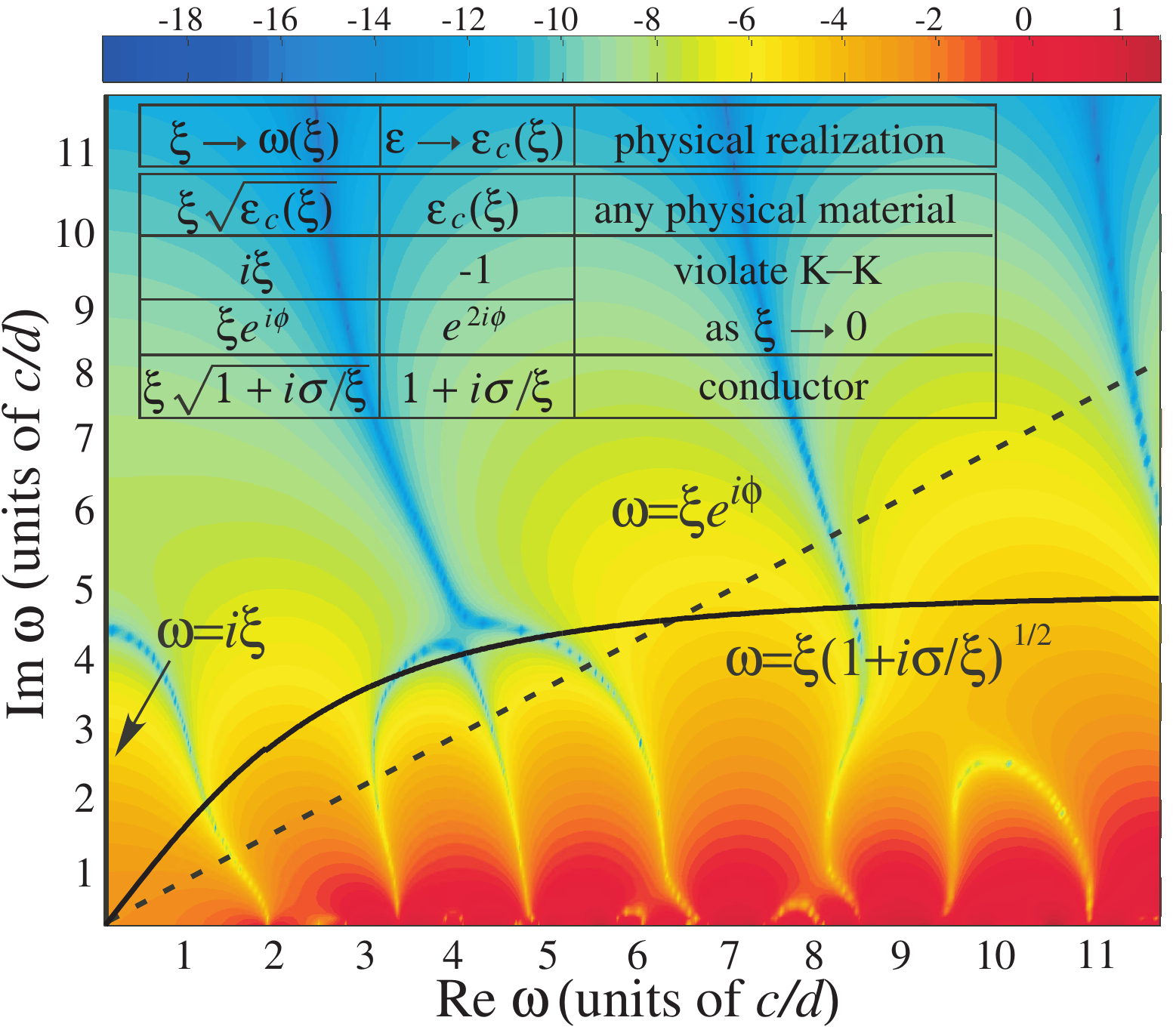}
\caption{Complex-frequency $\omega$ plot of the Casimir force
  integrand ($\ln |\Re dF_x / d\omega|$), where $dF_x / d\omega$ is in
  units of $\hbar/ d^2$, for the geometry of \figref{schematic}. As
  the real-$\omega$ axis is approached, the integrand becomes highly
  oscillatory, which is only partially revealed here due to the finite
  frequency resolution. Various integration contours of interest are
  labelled as black and dashed lines. (Inset:) Vacuum $\varepsilon=1$
  contour deformations $\omega(\xi)$ and their corresponding
  (real-frequency) physical realizations $\varepsilon_c(\xi)$.}
\label{fig:fvsw}
\end{figure}

However, the Wick rotation is not the only contour in the complex
plane that leads to a well-behaved decaying integrand. This is
illustrated by \figref{fvsw}, which shows the force integrand in the
complex plane for the piston-like geometry of \figref{schematic}.  In
particular, we calculate the ($x$-direction) force integrand
$dF_x/d\omega = \oiint_{\mathrm{surf}} \sum_j T_{xj}(\vec{r}, \omega)
dS_j$ on one square, for geometric parameters $h=0.5d$ and $s=d$,
where $d$ is the separation between the blocks. Here, we plot $\ln
|\Re dF_x/d\omega|$, which illustrates the basic features of the
integrand $d\omega/d\xi \, dF_x/d\omega$ ($\Im dF_x/d\omega$ is also
important, but is qualitatively similar).  As described above, this
integrand is oscillating along the $\Re \omega$ axis and decaying
along the $\Im \omega$ axis, but it also decays along any contour
where $\Im \omega$ is increasing (such as the three contours shown, to
be considered in more detail below).

The fact that $\omega$ and $\varepsilon$ only appear together in
\eqref{Green-gen}, as $\varepsilon \omega^2$, immediately suggests
that, instead of changing $\omega$ to a complex number, we can instead
operate at real frequencies by transforming $\varepsilon$. In
particular, a complex $\omega(\xi)$ is equivalent to a real frequency
$\xi$ with a complex $\varepsilon$, where the imaginary part of
$\varepsilon$ corresponds to dissipation loss.  Thus, an intuitive
explanation for why transforming to the complex-frequency plane was
numerically useful above is simply that it corresponds to lossy
materials that damp out the oscillations.  Given a medium
$\varepsilon$ at a complex frequency $\omega(\xi)$ where we wish to
compute the ST for the Casimir force, we will transform to an
equivalent problem at a real frequency $\xi$ and complex permittivity
$\varepsilon_\mathrm{c}$.  Namely, operating at a complex frequency
$\omega(\xi)$ is clearly equivalent (for the photon GF) to operating
at a real frequency $\xi$ and transforming a given material via
$\varepsilon(\xi) \to \varepsilon(\omega) \omega^2/\xi^2 =
\varepsilon_\mathrm{c}$.  Conversely, any frequency-dependent material
$\varepsilon_\mathrm{c}(\xi)$ at a given point in space can be related
to the GF for vacuum ($\varepsilon=1$) at that point by going from the
real frequency $\xi$ to a complex frequency $\omega = \xi
\sqrt{\varepsilon_\mathrm{c}(\xi)}$.

Because the ST is expressed in terms of the GF, and the GF at
microwave lengthscales is merely a rescaling of the GF at micron
lengthscales (if suitable materials can be found), one can conceivably
measure the GF in an experiment via the $S$-matrix elements of
antennas at centimeter scales, and so determine the Casimir force via
integration of the ST.  The passage to complex frequencies is
essential here, as in numerics, because measuring the real-frequency
GF will yield a highly oscillatory force integrand over an infinite
bandwidth, imposing significant experimental
challenges. Unfortunately, it is difficult to implement complex
frequency deformations directly, because a complex frequency
corresponds to fields and sources with exponential growth in time. An
alternative, suggested by the correspondence above, is to measure the
real-frequency GF of a physical medium with a complex permittivity
$\varepsilon_c(\xi) = \varepsilon(\omega) \omega^2 / \xi^2$.  This
$\varepsilon_c$ medium should satisfy two properties: it should
correspond to an $\omega$ contour where the ST integrand is rapidly
decaying, and it should be physically realizable. For simplicity, we
begin by considering complex contours for the ST in vacuum
($\varepsilon=1$).  The extension to arbitrary geometries/materials is
straight-forward: an arbitrary inhomogeneous medium
$\varepsilon(\vec{r},\xi)$ corresponds to $\varepsilon_c =
\omega^2\varepsilon(\vec{r},\omega) / \xi^2$.


If $\varepsilon_c(\xi)$ is to correspond to a physical medium, it must
satisfy the complex-conjugate property $\varepsilon_c(-\xi) =
\varepsilon_c(\xi)^*$ as well as the Kramers--Kronig (K--K)
relations~\cite{Jackson98}.  It is most important to satisfy these
conditions for small $\xi$, since the ST integrand is dominated by
long-wavelength contributions. One should also prohibit gain media,
which would lead to the exponentially growing fields we are trying to
avoid by not using complex $\omega$.  For example, Wick rotations
correspond to $\varepsilon_c(\xi) = -1$, and this is only possible at
$\xi=0$ in a gain medium, since in a dissipative medium,
$\varepsilon_c$ is real and positive along the whole imaginary-$\xi$
axis (this is implied by K--K).  The generalization to arbitrary
rotations in the complex plane $\varepsilon = e^{2i\phi}$ is both a
gain medium and violates $\varepsilon_c(-\xi) = \varepsilon_c(\xi)^*$
near $\xi=0$.  Thus, no realizable material can emulate these contours
even in a narrow bandwidth around $\xi=0$, as summarized on the table
(inset) of \figref{fvsw}.

Although traditional Wick rotations correspond to unphysical
materials, there are obviously many physical lossy materials to choose
from, each of which corresponds to a contour in the complex plane, and
one merely needs to find such a ``physical'' contour on which the ST
is rapidly decaying so that experiments can be performed over
reasonable bandwidths.  A simple and effective lossy material for this
purpose is a conductor with conductivity $\sigma$.  Because the
integral will turn out to be dominated by the contributions near zero
frequency, it is sufficient to consider $\sigma$ to be a constant (the
DC conductivity), although of course the full experimental
permittivity $\varepsilon_c(\xi)$ could also be used.  Specifically,
consider the general class of conductors defined by dispersion
relations of the form $\varepsilon_c(\xi) = 1 + i\sigma/\xi$,
corresponding to vacuum with a complex contour $\omega_\sigma =
\xi\sqrt{1 + i\sigma/\xi}$. As shown in \figref{fvsw}, the integrand
of this contour is in fact well behaved, rapidly decaying and exhibits
few oscillations.

\begin{figure}[tp]
\includegraphics[width=0.9\columnwidth]{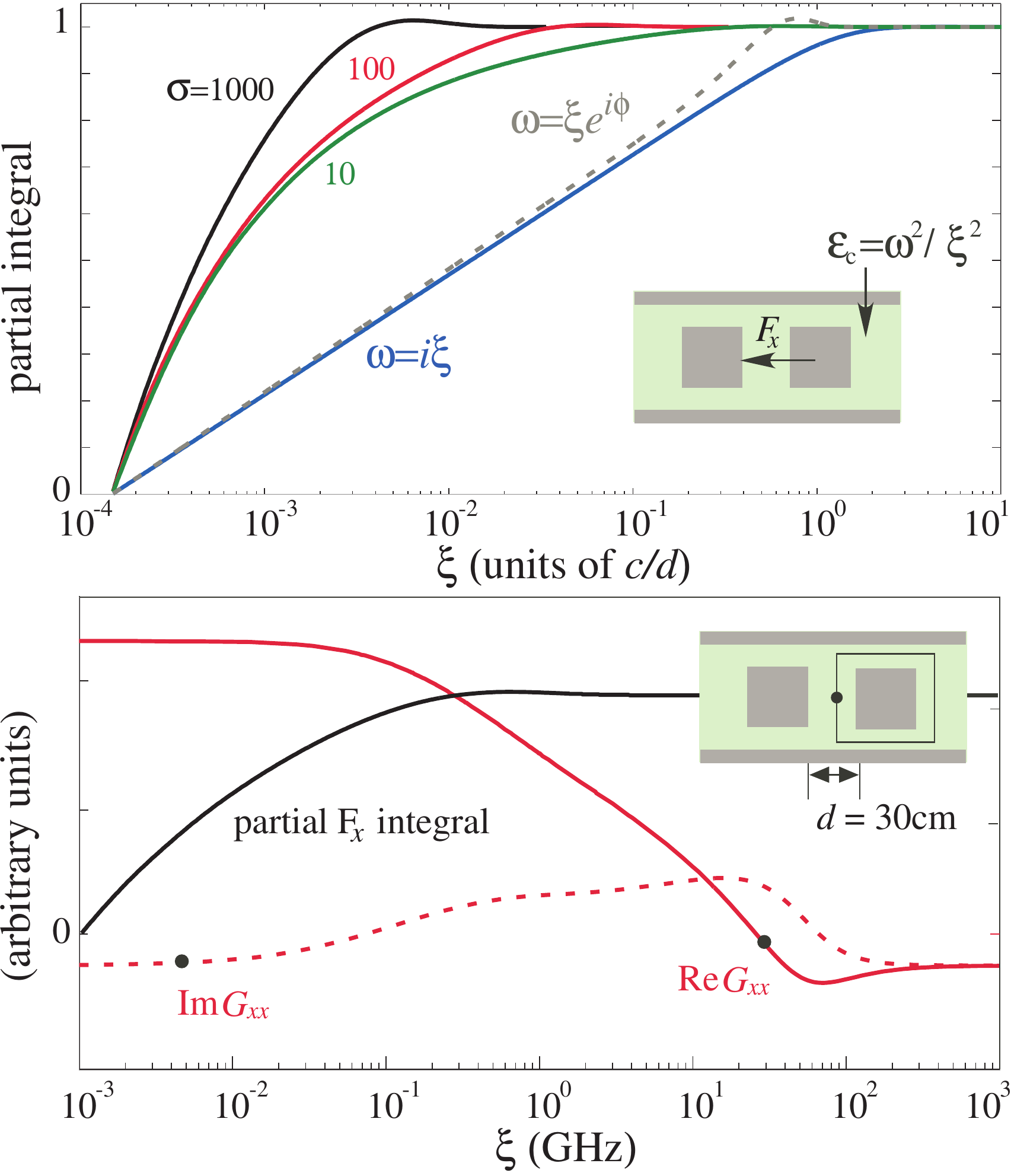}
\caption{(Top:) Partial force integral $\int_0^\xi dF_x$, normalized
  by $F_x$, as a function of $\xi$, for the various
  $\omega(\xi)$-contours (equivalently, various
  $\varepsilon_c=\omega^2/\xi^2$) shown in \figref{fvsxi}. The solid
  green, red and black lines correspond to conductive media with
  $\sigma= 10$, $10^2$ and $10^3$, respectively ($\sigma$ has units of
  $c/d$). The dashed grey and solid blue lines correspond to
  $\phi=\pi/4$ and $\phi=\pi/2$ (Wick) rotations. (Bottom:)
  Illustration of the required frequency bandwitdth for a possible
  realizations using saline solution at separation $d = 30$cm. The red
  lines plot the $xx$-component of the photon GF $G_{xx}$ at a single
  location on the surface contour (see inset) as a function of $\xi$
  (GHz). The black line is the corresponding partial force integrand.}
\label{fig:fvsxi}
\end{figure}

We now consider the Casimir force for the same structure as in
\figref{schematic}, still calculated by the same finite-difference
method as in \figref{fvsxi}, but we now focus on the properties along
different contour choices for both physical and unphysical media.  In
particular, \figref{fvsw}(top) plots the partial integral $\int_0^\xi
(d\omega/d\xi) (dF_x / d\omega) d\xi$, normalized by the total force
$\int_0^\infty dF_x$, as a function of $\xi$. [As it must, the total
  integral over $\xi$, the force $F_x$, is invariant regardless of the
  contour $\omega(\xi)$ and agrees with previous
  results~\cite{Rodriguez07:PRA}; Specifically, $F_x = 0.0335$ $(\hbar
  c / d^3)$.] We now comment on two important features of the
$\omega_\sigma$ contour that are relevant to experiments.

First, the Jacobian factor for $\omega_\sigma$ is given by
$d\omega_\sigma/d\xi = 0.5 (2 + i\sigma/\xi) / \sqrt{1 + i\sigma/\xi}$
and turns out to be very important at low $\xi$.  The ST integrand
itself goes to a constant as $\xi\to 0$ (due to the constant
contribution of zero-frequency modes), but the Jacobian factor
diverges in an integrable square-root singularity $\sim
\sqrt{\sigma/\xi}$.  Since this singularity is known analytically,
however, separate from the measured or calculated GF, integrating it
accurately poses no challenge.  Second, the larger the value of
$\sigma$, the more rapidly the ST integrand decays with $\xi$, and as
a consequence the force integral for larger $\sigma$ is dominated by
smaller $\xi$ contributions.  In comparison, previous calculations of
Casimir forces along the imaginary-frequency axis revealed that the
relevant $\xi$ bandwidth was determined by some characteristic
lengthscale of the geometry such as body
separations~\cite{Rodriguez07:PRA}.  Here, we have introduced a new
parameter $\sigma$ that can squeeze the relevant $\xi$ bandwidth into
a narrower region.  This ``spectral squeezing'' effect is potentially
useful for experiments, as it partially decouples the experimental
lengthscale of the geometry from the required frequency bandwidth.


As a consequence of the above results, we can now outline a possible
experiment at centimeter lengthscales that determines the Casimir
force at micron lengthscales, a Casimir analog computer (CAC). Suppose
that one wishes to compute the Casimir force between perfect-metal
objects separated by vacuum, such as the geometry in
\figref{schematic}.  One would then construct a scale model of this
geometry at a tabletop scale (e.g., centimeters) out of metallic
objects (which can be treated as perfect metals at microwave and
longer wavelengths).  To determine the ST integrand along a
complex-$\omega_\sigma$ contour, one would measure the GF at real
frequencies $\xi$ for the model immersed in a conducting fluid.  The
GF is related to the $S$-matrix of pairs of antennas, and the diagonal
of the GF to the $S$-matrix diagonal of a single antenna [noting that
  the finite size of the antenna automatically regularizes the
  integrand, as noted after \eqref{Green-gen}]. It is important for
the model structure to be large enough that the introduction of a
small dipole-like antenna does not significantly alter the
electromagnetic response. In general, the ST must be integrated in
space over a closed surface around the object, and correspondingly the
antenna's $S$-matrix spectrum must be measured at a number of antenna
positions (2d quadrature points) around this surface. (Unless one is
interested in computing the force on a single atom, which requires a
single antenna measurement.)  The different components of the GF
tensor correspond to different antenna orientations.  The magnetic GF
can be determined from the photon GF by \eqref{ecorr}, or possibly by
employing``magnetic dipole'' antennas formed by small current loops.

We now consider a particular CAC (at the cm scale) that employs
realistic geometric and material parameters. Many available fluids
exhibit almost exactly the desired material properties from above. One
such example is saline water, which has $\varepsilon(\xi) =
\varepsilon_s + i\sigma/\xi$, where $\varepsilon_s \approx 80$ and
$\sigma \approx 5$~S/m for relatively small values of salt
concentration~\cite{Klein77}. A calculation using these parameters,
based on the geometry of \figref{schematic}, assuming object sizes and
separations at the centimeter to meter scale (we choose $d = 0.3$~m
for the structure in \figref{schematic}, corresponding to a frequency
of 1~GHz), reveals that it is only required to integrate the stress
tensor up to small GHz frequencies $\xi$, which is well within the
reach of conventional antennas and electronics. This is illustrated
in~\figref{fvsxi}(bottom), which plots the $G_{xx}$ component of the
GF (red lines) as well as the partial force integrand (black line),
showing the high $\xi < 1$~GHz cancellations that occur once the ST is
integrated along a surface (inset).  We note that most salts exhibit
additional dispersion for $\xi > 10$~GHz~\cite{Klein77}, but we do not
need to reach those frequency scales. (Nevertheless, should there be
substantial dispersion in the conducting fluid, one could easily take
it into account as a different complex-$\omega$ contour.)

Some attention to detail is required in applying this correspondence
correctly.  For instance, using network analyzers, what is measured in
such an experiment is not the photon GF $\vec{G}$, but rather the
electric $S$-matrix $\vec{S}^E$ (the currents in a set of receiver
antennas due to currents in the source antennas) , related to the
electric GF (the $\vec{E}$-field response to an electric current
$\vec{J}$) by a factor depending on the antenna geometry alone
(relating $\vec{J}$ to $\vec{E}$). The electric GF will differ from
the photon GF by a factor of the real frequency $i\xi$. To summarize,
the photon GF will be given in terms of the measured $\vec{S}^E$ by
$G_{ij}(\omega) = (\alpha / i \xi) S^E_{ij}(\xi)$, where $\alpha$ is
the antenna-dependent geometric factor. To obtain the ST from
$G_{ij}$, one multiplies by factors of $\omega(\xi)^2$ as in
\eqref{ecorr}. \Figref{fvsxi}(bottom) illustrates the expected
behavior of $S_{xx} \sim G_{xx} / \xi$ in a realistic system employing
a sline solution with $d=30$cm.

The use of tabletop models and analog computers in physics, though
rarely explored in the context of vacuum fluctuations, continues to
play an important role in contemporary research fields, such as
quantum information~\cite{Lloyd08}. Especially for three-dimensional
geometries, tabletop experiments offer a route to rapidly exploring
many different geometric configurations that remain extremely
challenging for conventional numerical calculation.  Although many
details of such an experiment remain to be developed, we believe that
the basic ingredients are both clear and feasible, at least when
restricted to perfect-metal bodies.  The most difficult case to
realize seems to be the force between imperfect-metal or dielectric
bodies with a permittivity $\varepsilon(\omega)$, as the corresponding
tabletop system requires materials with a specified dispersion
relation $\varepsilon_c^\mathrm{body}(\xi) /
\varepsilon_c^\mathrm{fluid}(\xi) = \varepsilon(\omega(\xi))$ relative
to the conducting fluid.  This may be an opportunity for specially
designed meta-materials with the desired frequency response. 

We are grateful to Zheng Wang and Peter Shor at MIT, and to Jeremy
Munday at Caltech for useful discussions. This work was supported by
the Army Research Office through the ISN under Contract
No. W911NF-07-D-0004, the MIT Ferry Fund, and by US DOE Grant
No. DE-FG02-97ER25308 (ARW).


\end{document}